# Symmetry Adapted Residual Neural Network Diabatization: Conical Intersections in Aniline Photodissociation

Yifan Shen[a,*] and David R. Yarkony[b,*]

Department of Chemistry, Johns Hopkins University, Baltimore, Maryland 21218, USA

a) e-mail: yifanshensz@jhu.edu

b) e-mail: yarkony@jhu.edu

**ABSTRACT:** We present a symmetry adapted residual neural network (SAResNet) diabatization method to construct quasi-diabatic Hamiltonians that accurately represent *ab initio* adiabatic energies, energy gradients, and nonadiabatic couplings for moderate sized systems. Our symmetry adapted neural network inherits from the pioneering symmetry adapted polynomial and fundamental invariant neural network diabatization methods to exploit the power of neural network along with the transparent symmetry adaptation of polynomial for both symmetric and asymmetric irreducible representations. In addition, our symmetry adaptation provides a unified framework for symmetry adapted polynomial and symmetry adapted neural network, enabling the adoption of the residual neural network architecture, which is a powerful descendant of the pioneering feedforward neural network. Our SAResNet is applied to construct the full 36-dimensional coupled diabatic potential energy surfaces for aniline N-H bond photodissociation, with 2,269 data points and 32,640 trainable parameters and 190 $cm^{-1}$ root mean square deviation in energy. In addition to the experimentally observed $\pi\pi^*$ and $\pi$Rydberg/$\pi\sigma^*$ states, a higher state (HOMO - 1 $\pi$ to Rydberg/$\sigma^*$ excitation) is found to introduce an induced geometric phase effect thus indirectly participate in the photodissociation process.

## I. INTRODUCTION

When conical intersections[1–4] are involved in a nonadiabatic process[5–8], a quasi-diabatic[9,10] Hamiltonian ($\boldsymbol{H}^{\mathrm{d}}$) can greatly facilitate a reliable simulation. A variety of diabatizations have been reported in the literature, based on: smooth molecular properties,[11–16] ansatz of diabatic states,[17–20] configuration uniformity,[21–27] and nonadiabatic couplings.[28–31] When accuracy is the primary concern, it would be safest for diabatizations to employ multireference electronic structure methods which provide nonadiabatic couplings explicitly and analytically, such as multi-configurational self-consistent field[32–35] (MCSCF), extended multistate multireference second-order perturbation theory[36,37] (XMS-MRPT2), quasi-degenerate *N*-electron valence state second-order perturbation theory[38] (QD-NEVPT2), and multireference configuration interaction[39,40] (MRCI).

The focus of this work is the nonadiabatic-coupling-based regression method which accurately fits $\boldsymbol{H}^{\mathrm{d}}$ to energies, energy gradients, and nonadiabatic couplings obtained from *ab initio* electronic wave functions.[41–55] To exclude the distraction from external degrees of freedom, the regression is performed in internal rather than Cartesian coordinates. This internal coordinate system is not limited to bond lengths and angles and $3N$ - 6 degree of freedoms; instead, it can take any functional form and be redundant, so it is more appropriate to be considered as a feature extraction[56] from the raw Cartesian coordinates.

An important physical constraint $\boldsymbol{H}^{\mathrm{d}}$ must hold is the complete nuclear permutation inversion (CNPI) group symmetry. Originally, symmetry adapted polynomials (SAP) made from CNPI group symmetry adapted internal coordinates[57–60] are utilized to expand $\boldsymbol{H}^{\mathrm{d}}$ matrix elements.[41,48–50] Each CNPI group symmetry adapted internal coordinate carries an irreducible representation, so the polynomials made from them can have the resulting irreducibles conveniently determined by looking up the group multiplication table, no matter symmetric or asymmetric. As neural networks[56] (NN) become increasingly popular, fundamental invariant neural network[61–65] (FINN) is proposed, which consumes totally symmetric inputs to produce totally symmetric outputs. Due to the complication arising from NN layers and activations, antisymmetry is addressed by multiplying the FINN output with asymmetric modes.[44–46,51–53,55]

In this work, we revisit SAP and FINN to exploit the power of NN along with the transparent symmetry adaptation of polynomial for both symmetric and asymmetric irreducibles. In addition, our symmetry adaptation provides a unified framework for symmetry adapted polynomial and symmetry adapted neural network, enabling the adoption of the residual neural network[66] (ResNet) architecture, which is a powerful descendant of the pioneering feedforward NN. Section II elaborates our symmetry adapted residual neural network (SAResNet) approach, including the symmetry adaptation in section II.A, the residual connection in section II.B, and the diabatization protocol in section II.C. With an eye to a future dynamics simulation, our SAResNet diabatization is applied to construct full 36-dimensional coupled potential energy surfaces for aniline N-H bond photodissociation. The results are presented in section III. Section IV summarizes and discusses directions for future work.

## II. THEORY

### II.A Symmetry adapted neural network

Let us start with the mathematical form of NN. A standard *N*-layer feedforward NN is a sequence of *N* linear and non-linear transformations operated on the input vector $\boldsymbol{x}^0$ to obtain the output vector $\boldsymbol{y}^{N+1}$

$$\boldsymbol{y}^n = \boldsymbol{W}^n \boldsymbol{x}^{n-1} + \boldsymbol{b}^n, 1 \le n \le N+1 \tag{1a}$$

$$\boldsymbol{x}^n = \mathrm{A}(\boldsymbol{y}^n), 1 \le n \le N \tag{1b}$$

Where $n$ is the layer index (0 is the input layer, $N$+1 is the output layer, 1 to $N$ are the hidden layers), $\boldsymbol{x}$ is the neuron value vector, $\boldsymbol{W}$ is the linear combination weight matrix, $\boldsymbol{b}$ is the linear combination bias vector, $\boldsymbol{y}$ is the linearly transformed neuron value vector, A is the non-linear activation function.

To adapt to symmetry, we utilize SAP as $\boldsymbol{x}^0$, i.e., each vector element of $\boldsymbol{x}^0$ is an SAP term. For abelian groups, under symmetry operation $\hat{O}$ (e.g., in CNPI group symmetry, $\hat{O}$ can be the permutation of identical nuclei or the inversion of the entire molecule), $\boldsymbol{x}^0$ would preserve its absolute value, but its sign may or may not

$$\hat{O}\boldsymbol{x}^0 = \begin{cases} \boldsymbol{x}^0, \text{symmetric} \\ -\boldsymbol{x}^0, \text{asymmetric} \end{cases} \tag{2}$$

Non-abelian groups are beyond the scope of this paper. Although $\hat{O}$ is defined to operate on $\boldsymbol{x}^0$, it also operates on the consecutive $\boldsymbol{x}$s and $\boldsymbol{y}$s since they are derived from $\boldsymbol{x}^0$. $\boldsymbol{W}$ and $\boldsymbol{b}$ and A have no dependence on $\boldsymbol{x}^0$, so $\hat{O}$ has no effect on them and they commute with $\hat{O}$.

If we constrain $\boldsymbol{x}^0$ to be totally symmetric $\boldsymbol{x}_\mathrm{s}^0$, i.e., $\hat{O}\boldsymbol{x}_\mathrm{s}^0 = \boldsymbol{x}_\mathrm{s}^0$ for all possible $\hat{O}$, then we arrive at FINN

$$\hat{O}\boldsymbol{y}_\mathrm{s}^1 = \boldsymbol{W}_\mathrm{s}^1\hat{O}\boldsymbol{x}_\mathrm{s}^0 + \boldsymbol{b}_\mathrm{s}^1 = \boldsymbol{W}_\mathrm{s}^1\boldsymbol{x}_\mathrm{s}^0 + \boldsymbol{b}_\mathrm{s}^1 = \boldsymbol{y}_\mathrm{s}^1 \tag{3a}$$

$$\hat{O}\boldsymbol{x}_\mathrm{s}^1 = \mathrm{A}\left(\hat{O}\boldsymbol{y}_\mathrm{s}^1\right) = \mathrm{A}(\boldsymbol{y}_\mathrm{s}^1) = \boldsymbol{x}_\mathrm{s}^1 \tag{3b}$$

…

$$\hat{O}\boldsymbol{y}_\mathrm{s}^{N+1} = \boldsymbol{W}_\mathrm{s}^{N+1}\hat{O}\boldsymbol{x}_\mathrm{s}^N + \boldsymbol{b}_\mathrm{s}^{N+1} = \cdots = \boldsymbol{y}_\mathrm{s}^{N+1} \tag{3c}$$

Where subscript s is short for symmetric. What if we demand asymmetric $\boldsymbol{x}_\mathrm{a}^0$? In another word, there exists at least one $\hat{O}$ such that $\hat{O}\boldsymbol{x}_\mathrm{a}^0 = -\boldsymbol{x}_\mathrm{a}^0$

$$\hat{O}\boldsymbol{y}_\mathrm{a}^1 = \boldsymbol{W}_\mathrm{a}^1\hat{O}\boldsymbol{x}_\mathrm{a}^0 + \boldsymbol{b}_\mathrm{a}^1 = -\boldsymbol{W}_\mathrm{a}^1\boldsymbol{x}_\mathrm{a}^0 + \boldsymbol{b}_\mathrm{a}^1 \tag{4a}$$

$$\hat{O}\boldsymbol{x}_\mathrm{a}^1 = \mathrm{A}\left(\hat{O}\boldsymbol{y}_\mathrm{a}^1\right) \tag{4b}$$

Where subscript a is short for asymmetric. As we can see from Equation (4a), $\hat{O}\boldsymbol{y}_\mathrm{a}^1$ does not necessarily equal to $-\boldsymbol{y}_\mathrm{a}^1$, unless $\boldsymbol{b}_\mathrm{a}^1 = 0$

$$\boldsymbol{y}_\mathrm{a}^1 = \boldsymbol{W}_\mathrm{a}^1\boldsymbol{x}_\mathrm{a}^0 + \boldsymbol{b}_\mathrm{a}^1 = \boldsymbol{W}_\mathrm{a}^1\boldsymbol{x}_\mathrm{a}^0 \tag{5a}$$

$$\hat{O}\boldsymbol{y}_\mathrm{a}^1 = \boldsymbol{W}_\mathrm{a}^1\hat{O}\boldsymbol{x}_\mathrm{a}^0 = -\boldsymbol{W}_\mathrm{a}^1\boldsymbol{x}_\mathrm{a}^0 = -\boldsymbol{y}_\mathrm{a}^1 \tag{5b}$$

Plug Equation (5) into Equation (4b)

$$\hat{O}\boldsymbol{x}_{\mathrm{a}}^{1} = \mathrm{A}\left(\hat{O}\boldsymbol{y}_{\mathrm{a}}^{1}\right) = \mathrm{A}(-\boldsymbol{y}_{\mathrm{a}}^{1}) \tag{6}$$

Equation (6) indicates that to have $\hat{O}\boldsymbol{x}_{\mathrm{a}}^{1} = -\boldsymbol{x}_{\mathrm{a}}^{1}$, A must be an odd function

$$\hat{O}\boldsymbol{x}_{\mathrm{a}}^{1} = \mathrm{A}_{\mathrm{odd}}(-\boldsymbol{y}_{\mathrm{a}}^{1}) = -\,\mathrm{A}_{\mathrm{odd}}(\boldsymbol{y}_{\mathrm{a}}^{1}) = -\boldsymbol{x}_{\mathrm{a}}^{1} \tag{7}$$

By induction, now the sign flip gets carried all the way down to the output layer as we wish

$$\hat{O}\boldsymbol{y}_{\mathrm{a}}^{N+1} = \boldsymbol{W}_{\mathrm{a}}^{N+1}\hat{O}\boldsymbol{x}_{\mathrm{a}}^{N} = \cdots = -\boldsymbol{y}_{\mathrm{s}}^{N+1} \tag{8}$$

In summary, the mathematical form of symmetry adapted neural networks (SANN) for the asymmetric irreducibles is

$$\boldsymbol{y}_{\mathrm{a}}^{n} = \boldsymbol{W}_{\mathrm{a}}^{n}\boldsymbol{x}_{\mathrm{a}}^{n-1}, 1 \le n \le N+1 \tag{9a}$$

$$\boldsymbol{x}_{\mathrm{a}}^{n} = \mathrm{A}_{\mathrm{odd}}(\boldsymbol{y}_{\mathrm{a}}^{n})\,, 1 \le n \le N \tag{9b}$$

Among the conventional activation functions, the hyperbolic tangent (tanh) is an odd function, so we would stick to $\mathrm{A}_{\mathrm{odd}}$ = tanh in this work. For the totally symmetric irreducible, SANN does not have to be different from FINN

$$\boldsymbol{y}_{\mathrm{s}}^{n} = \boldsymbol{W}_{\mathrm{s}}^{n}\boldsymbol{x}_{\mathrm{s}}^{n-1} + \boldsymbol{b}_{\mathrm{s}}^{n}, 1 \le n \le N+1 \tag{10a}$$

$$\boldsymbol{x}_{\mathrm{s}}^{n} = \mathrm{A}(\boldsymbol{y}_{\mathrm{s}}^{n})\,, 1 \le n \le N \tag{10b}$$

Although we possess the flexibility to adopt arbitrary activation function, since tanh is still a legitimate choice, we would also stick to A = tanh in this work. Xavier initialization[67] is then applied in accordance with tanh.

### II.B Residual connection

One desirable property of SANN is that SANN regression is a natural extension of SAP regression, which can now be considered as a 0-layer SANN regressor.

$$\boldsymbol{y}_{\mathrm{a}}^{1} = \boldsymbol{W}_{\mathrm{a}}^{1}\boldsymbol{x}_{\mathrm{a}}^{0} \tag{11a}$$

$$\boldsymbol{y}_{\mathrm{s}}^{1} = \boldsymbol{W}_{\mathrm{s}}^{1}\boldsymbol{x}_{\mathrm{s}}^{0} + \boldsymbol{b}_{\mathrm{s}}^{1} \tag{11b}$$

This unification of SAP and SANN allows us to benefit from the success of both polynomial and neural network diabatizations by combining the SAP and the SANN outputs together

$$\boldsymbol{y}_{\mathrm{a}}^{n} = \boldsymbol{W}_{\mathrm{a}}^{n}\boldsymbol{x}_{\mathrm{a}}^{n-1}, 1 \leq n \leq N \tag{12a}$$

$$\boldsymbol{x}_{\mathrm{a}}^{n} = \mathrm{A}_{\mathrm{odd}}(\boldsymbol{y}_{\mathrm{a}}^{n}), 1 \leq n \leq N \tag{12b}$$

$$\boldsymbol{y}_{\mathrm{a}}^{N+1} = \boldsymbol{W}_{\mathrm{a}}^{N+1}\boldsymbol{x}_{\mathrm{a}}^{N} + \boldsymbol{W}_{\mathrm{a}}^{\mathrm{res}}\boldsymbol{x}_{\mathrm{a}}^{0} \tag{12c}$$

$$\boldsymbol{y}_{\mathrm{s}}^{n} = \boldsymbol{W}_{\mathrm{s}}^{n}\boldsymbol{x}_{\mathrm{s}}^{n-1} + \boldsymbol{b}_{\mathrm{s}}^{n}, 1 \leq n \leq N \tag{12d}$$

$$\boldsymbol{x}_{\mathrm{s}}^{n} = \mathrm{A}(\boldsymbol{y}_{\mathrm{s}}^{n}), 1 \leq n \leq N \tag{12e}$$

$$\boldsymbol{y}_{\mathrm{s}}^{N+1} = \boldsymbol{W}_{\mathrm{s}}^{N+1}\boldsymbol{x}_{\mathrm{s}}^{N} + \boldsymbol{b}_{\mathrm{s}}^{N+1} + \boldsymbol{W}_{\mathrm{s}}^{\mathrm{res}}\boldsymbol{x}_{\mathrm{s}}^{0} \tag{12f}$$

Computer scientifically, this combination of the SAP and the SANN outputs is a residual connection, which upgrades the simple feedforward NN to the ResNet, so our SANN also gets upgraded to SAResNet. ResNet offers many benefits, including the avoidance of vanishing gradients, the mitigation of degradation, etc. Chemically, SAResNet reflects the idea of utilizing low-order polynomial for qualitative correctness and complicated functions (such as NN) for quantitative correction. One illustrative scenario is the bound coordinates. Consider a bound internal mode, where we can have an upward parabola (quadratic polynomial) for qualitative description: potential energy should be low around the equilibrium, but go up fast if moving too far away. Incorporation of higher-order polynomials would make quantitative correction around the equilibrium, but may destroy the qualitative correctness when moving further away, due to its oscillatory behavior. NN would not fluctuate wildly, since the supremum for $|y^{N+1}|$ is $||\boldsymbol{W}^{N+1}||_1$, which forbids NN from creating the high-energy wall alone but makes NN a perfect choice for quantitative correction as long as an appropriate regularization[68] is posed on $\boldsymbol{W}^{N+1}$. For the dissociative coordinates, a quadratic polynomial of exponential functions (a.k.a. Morse potential) would produce a qualitative potential energy curve, then NN would polish it to perfection.

### II.C Diabatization protocol

The complete $\boldsymbol{H}^{\mathrm{d}}$ model includes:

1. Feature extraction from Cartesian coordinate: convert Cartesian coordinates to CNPI group symmetry adapted internal coordinates.
2. Polynomial features: multiply CNPI group symmetry adapted internal coordinates to obtain SAPs.
3. SAResNet: instantiate one SAResNet for each $\boldsymbol{H}^{\mathrm{d}}$ matrix element, then according to the $\boldsymbol{H}^{\mathrm{d}}$ matrix element irreducible use the same-irreducible SAP as the input layer, finally process the input SAP with Equations (12a) to (12c) if asymmetric or (12d) to (12f) if symmetric.

4. $\boldsymbol{H}^{\mathrm{d}}$ matrix: collect all SAResNet outputs into one real symmetric matrix.

The $\boldsymbol{H}^{\mathrm{d}}$ model architecture is sketched in Figure 1.

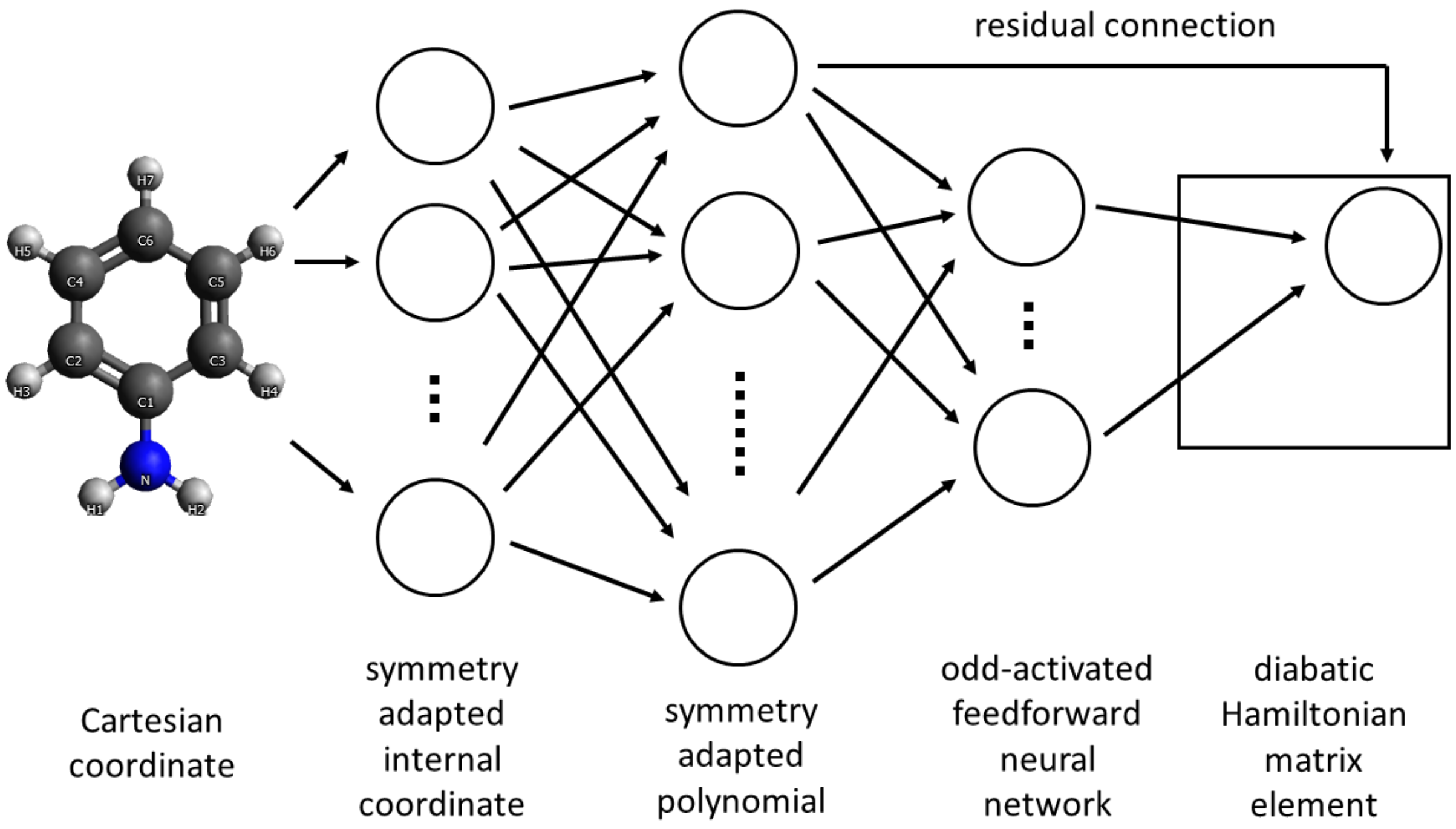

**Figure 1.** The symmetry adapted residual neural network diabatic Hamiltonian model architecture. The protocol to obtain the upper right matrix element is sketched. Every other matrix element follows the same protocol.

Once we have the $\boldsymbol{H}^{\mathrm{d}}$ model architecture set up, the next step is to train the $\boldsymbol{H}^{\mathrm{d}}$ model parameters (linear combination weights and biases). Given a training set gathered from *ab initio* electronic structure computation, we minimize the loss function $L(\boldsymbol{c})$[54]

$$L(\boldsymbol{c}) = \sum_{i=1}^{M} w_i \left[\rho^2 \left\|\boldsymbol{H}_r^{\mathrm{d}}(\boldsymbol{R}_i;\boldsymbol{c}) - \boldsymbol{H}_r^{ab}(\boldsymbol{R}_i)\right\|_F^2 + \left\|\nabla\boldsymbol{H}_r^{\mathrm{d}}(\boldsymbol{R}_i;\boldsymbol{c}) - \nabla\boldsymbol{H}_r^{ab}(\boldsymbol{R}_i)\right\|_F^2\right] + \mu\left\|\boldsymbol{c} - \boldsymbol{c}_{\mathrm{p}}\right\|_2^2 \quad (13)$$

Where $\boldsymbol{c}$ is the trainable parameter vector, $M$ is the number of geometries contained in the training set, $w$ is the fitting weight, $\rho$ is a scaling factor accounting for the unit difference[69] between $\boldsymbol{H}$ and $\nabla\boldsymbol{H}$, superscript d and *ab* indicate coming from $\boldsymbol{H}^{\mathrm{d}}$ or *ab initio*, subscript $r$ denotes the matrix representation of the operator (composite for quasi-degenerate data points, adiabatic for others) and $F$ denotes the Frobenius norm, $\mu$ is the regularization strength, $\boldsymbol{c}_{\mathrm{p}}$ is the prior estimation of $\boldsymbol{c}$. Trust region[70] and Dai-Yuan conjugate gradient[71] algorithms are applied to minimize $L(\boldsymbol{c})$. More details can be found in supporting information.

Before we end the theory section, we would like to comment on the role of nonadiabatic coupling in our diabatization. Training with Equation (13) is how we utilize nonadiabatic coupling. In adiabatic representation, $\boldsymbol{H}$ is a diagonal matrix with energies as its diagonal elements, the diagonal elements of $\nabla\boldsymbol{H}$ are energy gradients by Hellmann-Feynman theorem, and the off-diagonal elements of $\nabla\boldsymbol{H}$ are energy-difference-scaled nonadiabatic couplings

$$[\nabla\boldsymbol{H}_{\mathrm{adiabatic}}]_{ij} = \left(E_j - E_i\right) * d_{ij} \tag{14}$$

Where $E_i$ is the electronic energy of adiabatic electronic state $i$, $d_{ij}$ is the nonadiabatic coupling between adiabatic electronic states $i$ and $j$. Through minimizing the least square difference from *ab initio* $\boldsymbol{H}$ and $\nabla\boldsymbol{H}$, our model $\boldsymbol{H}^{\mathrm{d}}$ is the optimal diabatization to the *ab initio* Hamiltonian in terms of the square error to energies, energy gradients, and nonadiabatic couplings. We would like to point out that, under the least square framework, nonadiabatic coupling or other interstate information is indispensable during diabatization process, because otherwise the adiabatic potential energy surfaces would be a perfect fit to energies and energy gradients, even though they are in no way diabatic. Once the diabatization is done, however, nonadiabatic coupling is no longer necessary, and the nonadiabatic dynamics could be carried out in diabatic representation solely with our model $\boldsymbol{H}^{\mathrm{d}}$.

## III. RESULTS AND DISCUSSION

Aromatic amines are present in many biological molecules such as the DNA bases, so their spectroscopy and photochemistry are of great interest to understand and treat UV-induced mutagenesis and carcinogenesis.[72,73] As the prototypical aromatic amine, aniline ($C_6H_5NH_2$) serves as a steppingstone to greater understanding of the photodynamics of more complex aromatic amines. The lowest-energy photodissociation channel in aniline is the N-H bond breaking:

$$C_6H_5NH_2 + h\upsilon \rightarrow C_6H_5NH + H \tag{15}$$

The other dissociation channels, e.g. the C-N bond breaking, are at least 2 eV higher,[74] consuming at least 50% more energy than the N-H bond breaking, so they are beyond the scope of this paper.

Based on experiments[74–79] and computations,[80–83] for the lowest and fastest dissociation there are 3 singlet states involved: the closed-shell ground state ($S_0$), a $\pi\pi^*$ state ($S_1$), and a $\pi$Rydberg/$\pi\sigma^*$ state ($S_2$). Since the highest possible point group aniline can possess is $C_{2v}$, we make symmetry arguments with the $G_4$ subgroup of the CNPI group, which is isomorphic to $C_{2v}$. Therefore, we can label these states of interest more clearly by $G_4$ irreducibles: $S_0$ is ${}^1A_1$, $S_2$ is ${}^1B_1$, $S_1$ is ${}^1B_2$. A ${}^1A_2$ state (HOMO - 1 $\pi$ to Rydberg/$\sigma^*$ excitation) is additionally computed since with ${}^1B_2$ it forms a conical intersection, which is then found to introduce an induced geometric phase effect.

## III.A Electronic structure

We collect *ab initio* electronic structure data from a multireference configuration interaction with single and double excitation (MRCISD) description, with 18 frozen core orbitals and a 14-electron 14-orbital restricted active space up to double excitation. The active space consists of the 7 $\Pi_7^8$ orbitals, a 3s Rydberg orbital, the 2 N-H σ and $\sigma^*$ pairs, and the C-N σ and $\sigma^*$ pair. The 2-external excitations from the lowest two active orbitals are excluded. This MRCISD is comprised of 158,587,185 configuration state functions.

A standard atomic orbital basis set (cc-pVTZ[84] for the nitrogen and the carbons, cc-pVDZ[84] for the hydrogens) is sufficient to describe the valence orbitals, but the 3s Rydberg orbital requires a special diffuse function. To generate it, following Roos *et al.*[85] we place a set of diffusive Gaussian functions on the nitrogen, run complete active space self-consistent field on the self-consistent field optimized $^1B_1$ state minimum, then take the linear combination coefficients. In addition, due to the linear-dependency issue, we shrink the outmost p orbital of nitrogen and s orbital of carbon by a factor of 3. The customized atomic orbital basis set and other details are reported in supporting information. All electronic structure computations are performed using the COLUMBUS[39,40,86–92] suite of programs.

## III.B Construction of $H^d$

Although $G_4$ is enough to interpret the electronic structure data, it misses out the amino ($NH_2$) group rotation, which requires a $G_8$ group to capture. The fact that all interesting electronic states doubly occupy the N-H σ orbitals in the bound region indicates that they are all symmetric with respect to the $NH_2$ hydrogen permutation, so we choose to let the diabatic electronic states carry the g subset of $G_8$ irreducibles: $^1A_{1g}$, $^1B_{1g}$, $^1B_{2g}$, $^1A_{2g}$. The detailed group theory analysis can be found in supporting information.

We define 63 CNPI group symmetry adapted internal coordinates, then mix them into 944 $^1A_{1g}$ SAPs, 324 $^1B_{1g}$ SAPs, 383 $^1B_{2g}$ SAPs, 299 $^1A_{2g}$ SAPs. The details can be found in supporting information. We use 1-layer SAResNet with 5 hidden neurons for $\boldsymbol{H}^d$ diagonals, and 1-layer SANN with 8 hidden neurons for off-diagonals. There are 32,640 trainable parameters, which are determined on a training set with 2,269 data points (corresponding to 825,916 least square equations). Our root mean square deviation of energy is 190 $cm^{-1}$. The satisfying capability to achieve adequate accuracy with affordable model size and training set size has demonstrated the promising potential of SAResNet diabatization in dealing with even larger systems.

**III.C Photodissociation mechanism**

Starting from equilibrium, ultraviolet photon vertically and diabatically excites aniline from $^1A_1$ to $^1B_2$, since the transition dipole moment from $^1A_1$ to $^1B_2$ is 20 times larger than the one from $^1A_1$ to $^1B_1$. Through internal conversion facilitated by the $^1B_1$-$^1B_2$ conical intersection, the electronic wave function propagates from $^1B_2$ to $^1B_1$, which would then lead to dissociation. The $^1B_1$ state is initially bound but would turn dissociative after a saddle point, due to the character switch from Rydberg to N-H $\sigma^*$. A $^1A_1$-$^1B_1$ conical intersection sits between the saddle point and the dissociation limit, branching the wave packet into a fast direct $^1B_1$ dissociation and a $^1A_1$ rebound back to the bound region. Although not observed in experiment since not directly involved in the dissociation dynamics, a $^1B_2$-$^1A_2$ conical intersection exists between the saddle point and the $^1A_1$-$^1B_1$ conical intersection. Figure 2 reports these critical geometries in the order of appearance during the photodissociation process.

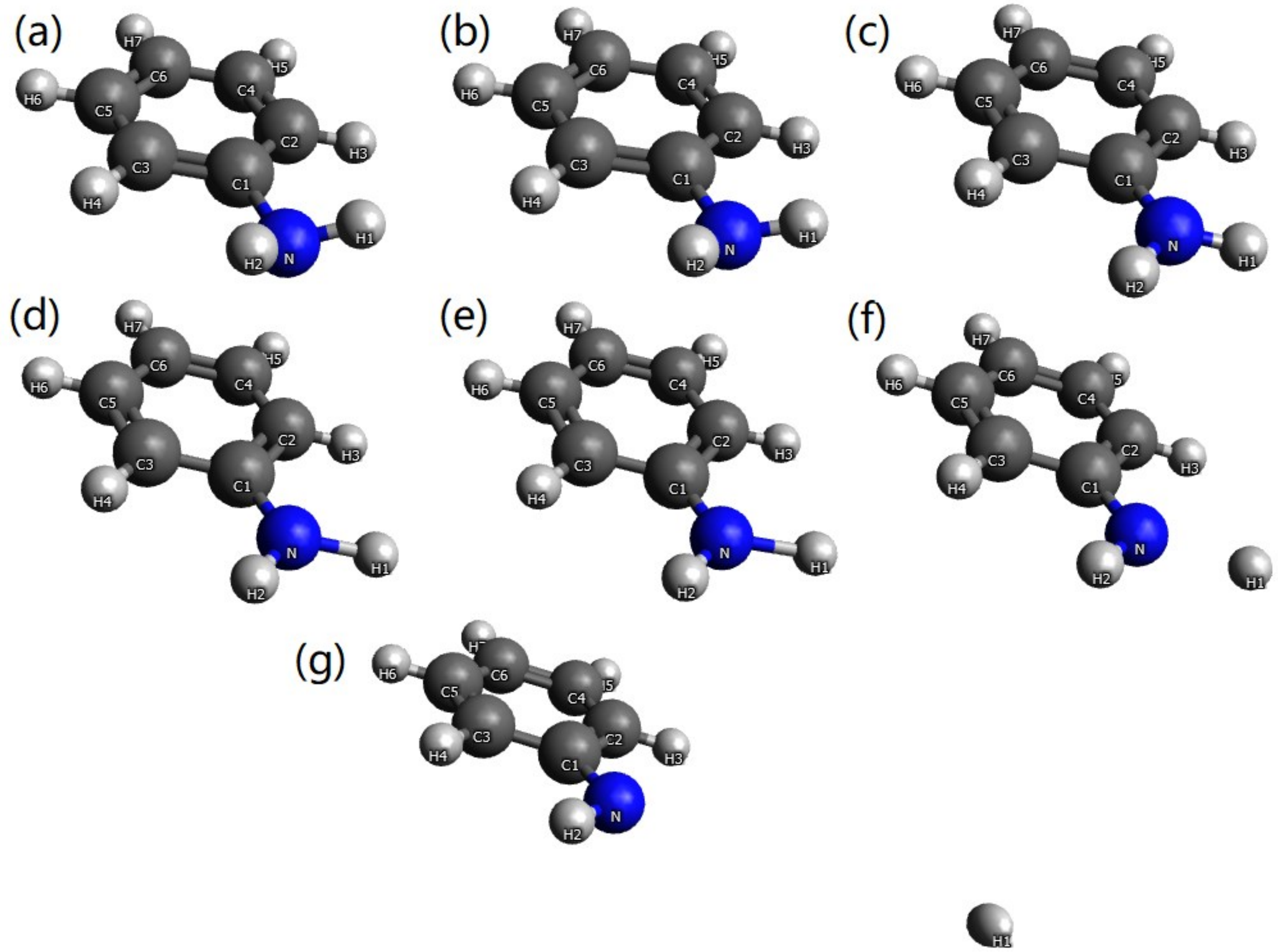

**Figure 2.** The critical geometries encountered during aniline photodissociation: (a) ground state minimum (min-$^1A_1$); (b) $\pi\pi^*$-$\pi$Rydberg minimum energy crossing (mex-$^1B_1$-$^1B_2$); (c) Rydberg

state minimum (min-$^1B_1$); (d) dissociation saddle point (sad-$^1B_1$); (e) $^1B_2$-$^1A_2$ minimum energy crossing (mex-$^1B_2$-$^1A_2$); (f) $^1A_1$-$^1B_1$ minimum energy crossing (mex-$^1A_1$-$^1B_1$); (g) $\pi\sigma^*$ dissociation limit (asymptote).

As we can see from Figure 2a to 2c, the $^1A_1$ state minimum (min-$^1A_1$), $^1B_1$-$^1B_2$ minimum energy crossing (mex-$^1B_1$-$^1B_2$), $^1B_1$ state minimum (min-$^1B_1$) have their N-H bonds bound but the amino group rocks from considerable phenyl-amino dihedral angle (≈ C1 out of $NH_2$ plane angle) to none. By inertia, in dynamics the amino group would rock to the other side of the benzene plane, visiting the symmetry-related counterparts of min-$^1A_1$ and mex-$^1B_1$-$^1B_2$. From Figure 2d to 2g, the amino group is now fixed to be within benzene plane; it is the N-H bond stretching that links min-$^1B_1$, the $^1B_1$ state dissociation saddle point (sad-$^1B_1$), $^1B_2$-$^1A_2$ minimum energy crossing (mex-$^1B_2$-$^1A_2$), $^1A_1$-$^1B_1$ minimum energy crossing (mex-$^1A_1$-$^1B_1$), and the $^1B_1$ state dissociation limit (asymptote), as in usual dissociative processes.

An ensemble of nonadiabatic trajectories say the photodissociation mechanism best. We run fewest switches surface hopping[93] based on our constructed $\boldsymbol{H}^d$ and illustrate a representative trajectory in Figure 3. As we can see from Figure 3, the trajectory starts at min-$^1A_1$, with an ultraviolet photon electronically exciting aniline from $^1A_1$ to $^1B_2$. Following the potential energy curvature, aniline goes from min-$^1A_1$ to mex-$^1B_1$-$^1B_2$, where it circles around the dual mex-$^1B_1$-$^1B_2$ for 40 fs to transfer electronic population from $^1B_2$ to $^1B_1$. After the $^1B_2$ to $^1B_1$ transition, aniline would stay close to mex-$^1B_1$-$^1B_2$ for another 200 fs to accumulate kinetic energy along the N-H stretching. Finally, with ample kinetic energy, aniline breaks through sad-$^1B_1$ and marches toward the dissociation limit. In this trajectory, aniline would stay on the $^1B_1$ state when passing through mex-$^1A_1$-$^1B_1$ and reach the dissociation limit, which only takes 20 fs from sad-$^1B_1$ to the asymptote. In other trajectories, aniline may undergo another internal conversion to fall back to the $^1A_1$ state, which is a bound state and would bring aniline back to the bound region.

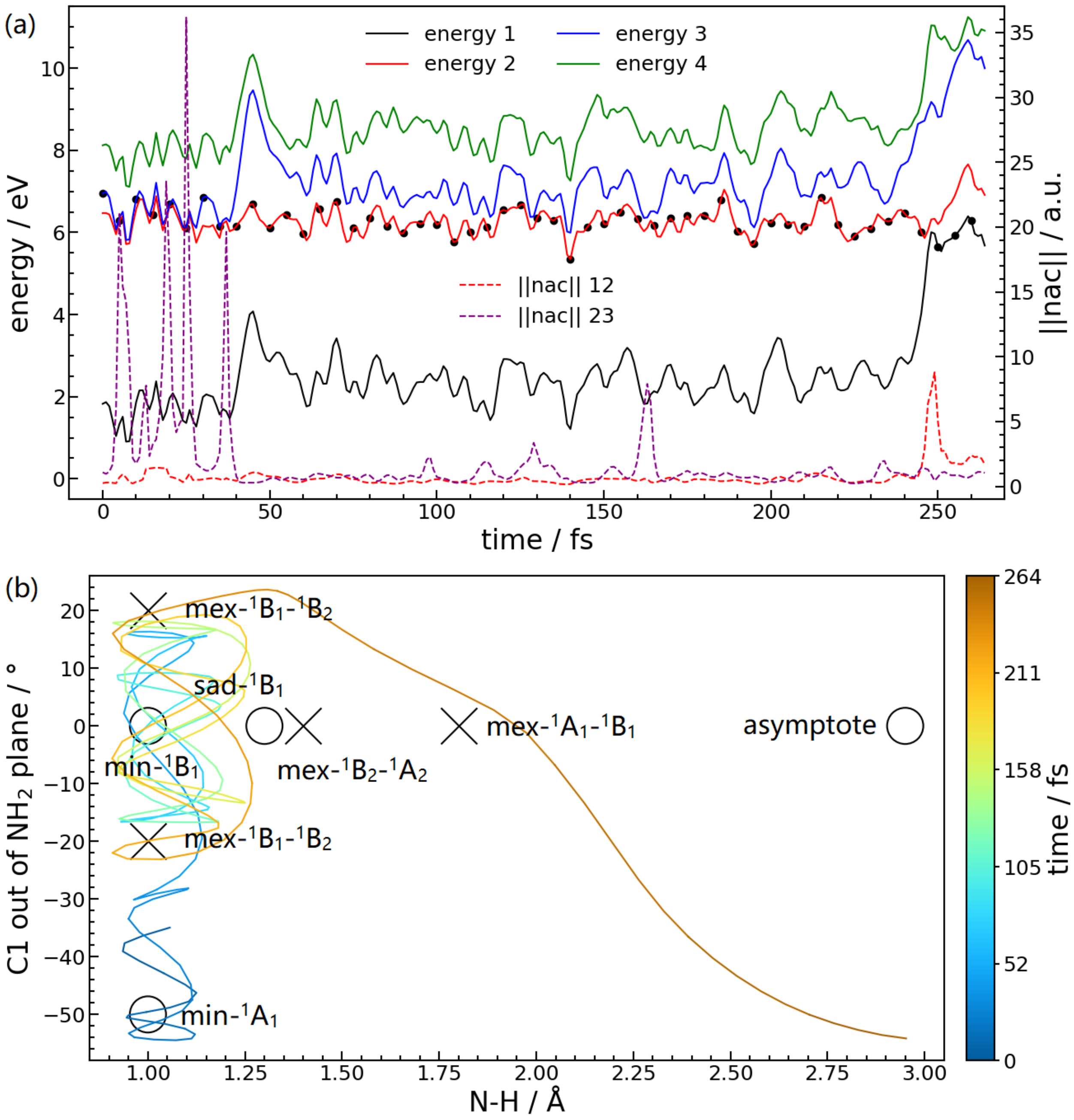

**Figure 3.** A representative surface hopping trajectory. (a) The potential energies and norm of nonadiabatic couplings (||nac||) with respect to time. Black dots indicate the active state. (b) The movement in the C1 out of $NH_2$ plane angle – N-H bond length space over time. Locations of critical geometries are approximately marked.

We investigate the potential energy curves along the two key internal coordinates: the C1 out of $NH_2$ plane angle and the N-H bond length. The linear synchronous transit path connecting the dual mex-$^1B_1$-$^1B_2$ is presented in Figure 4. $\boldsymbol{H}^d$ reproduces MRCISD energies and nonadiabatic couplings quantitatively.

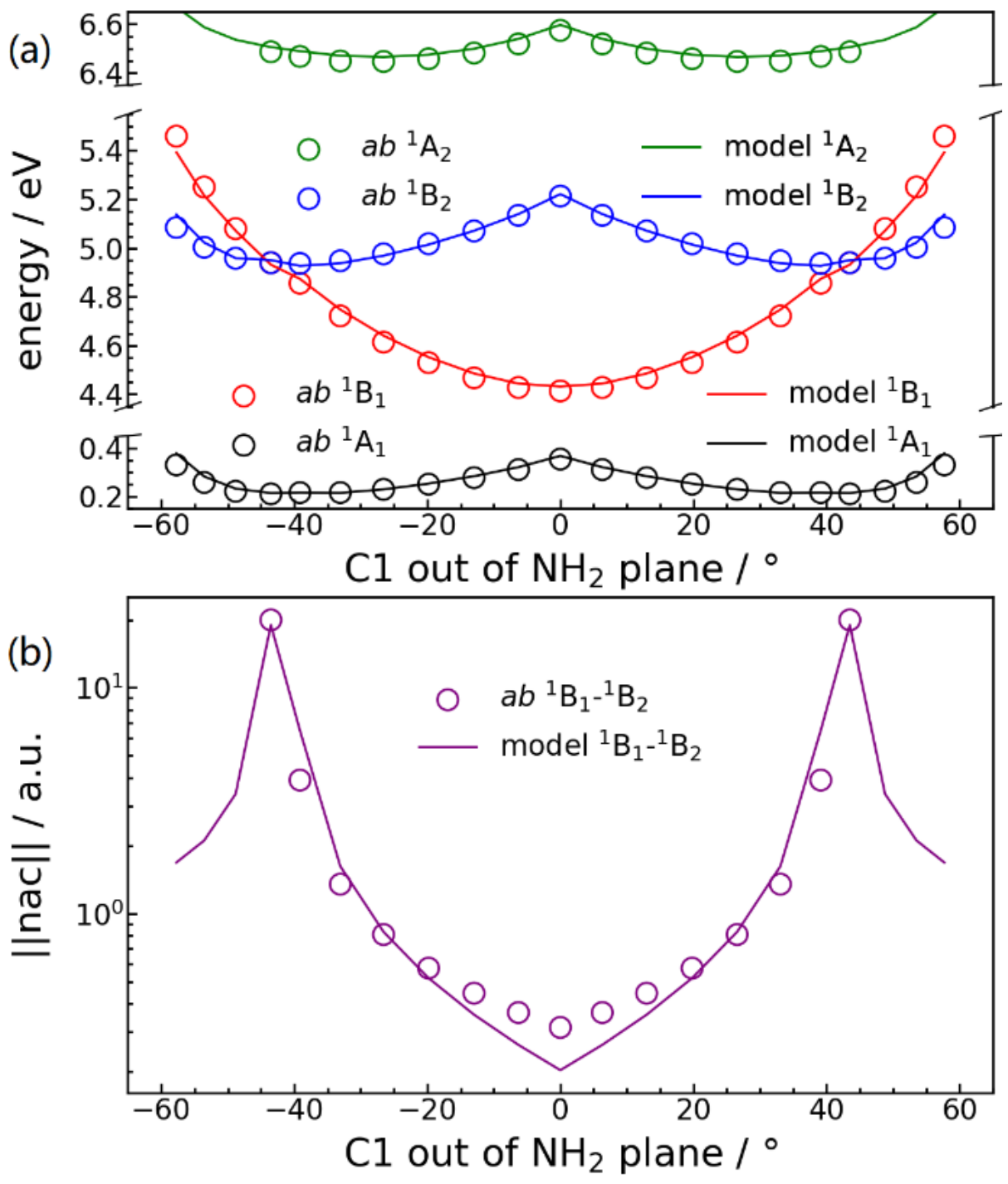

**Figure 4.** The linear synchronous transit path connecting dual $^1B_1$-$^1B_2$ minimum energy conical intersections. (a) The potential energies. (b) The norm of nonadiabatic couplings (||nac||) between $^1B_1$ and $^1B_2$ states. The infinite nonadiabatic couplings at $^1B_1$-$^1B_2$ conical intersections are arbitrarily set to 20 a.u. for plot purpose.

The $^1B_1$ energy optimized path from min-$^1B_1$ to asymptote can be found in Figure 5. Again, $\boldsymbol{H}^d$ reproduces MRCISD energies quantitatively. The $^1B_2$-$^1A_2$ and the $^1A_1$-$^1B_1$ conical intersections emerge along this path, due to the bound nature of the $^1A_1$ and the $^1B_2$ states and the dissociative character of the $^1B_1$ and the $^1A_2$ states. The dissociation barrier height on the $^1B_1$ state surface is 0.5 eV.

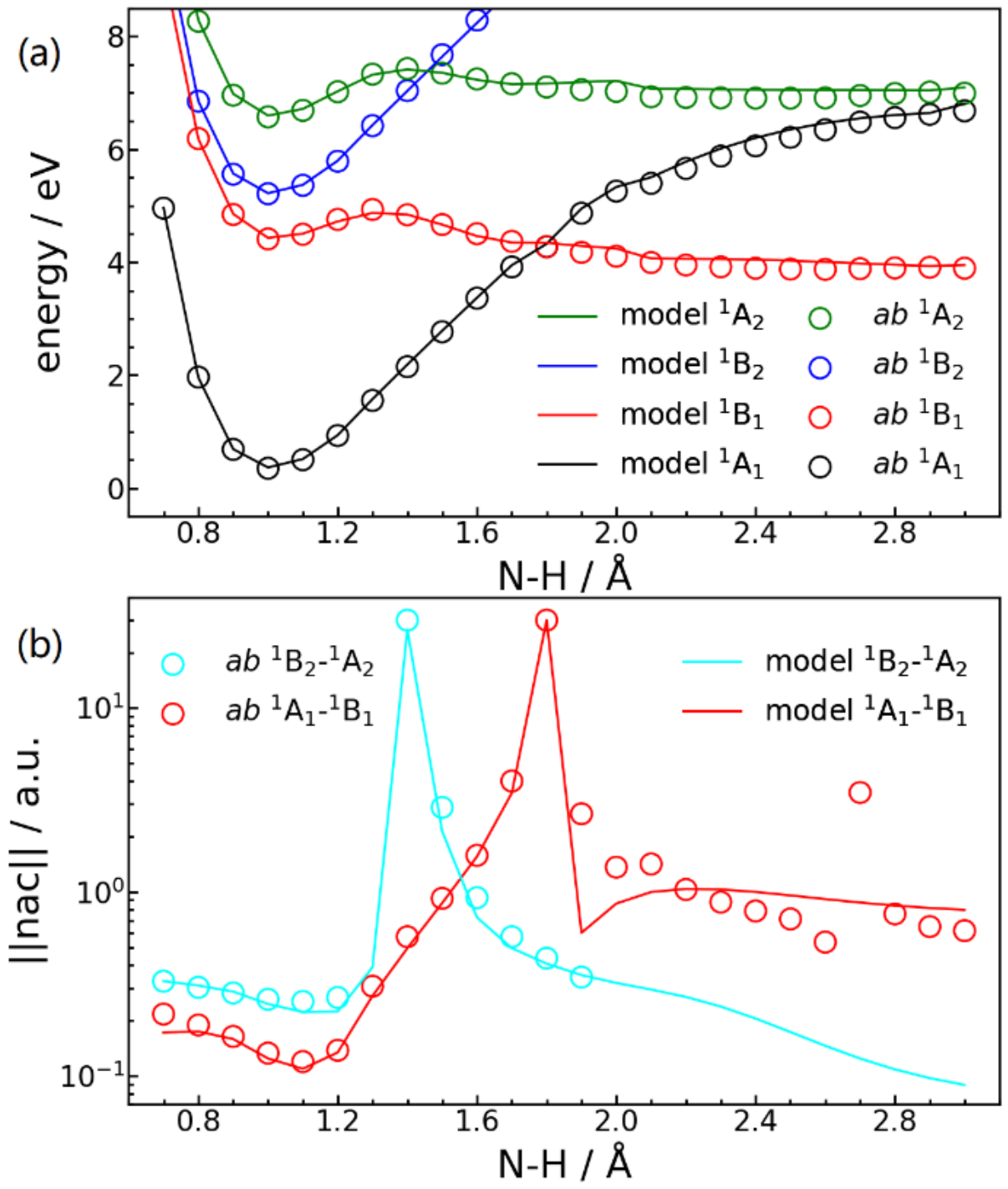

**Figure 5.** The $^1B_1$ energy optimized path from $^1B_1$ minimum to dissociation limit. (a) The potential energies. (b) The norm of nonadiabatic couplings (||nac||) between $^1A_1$ and $^1B_1$ states and between $^1B_2$ and $^1A_2$ states. The infinite nonadiabatic couplings at $^1B_2$-$^1A_2$ and $^1A_1$-$^1B_1$ conical intersections are arbitrarily set to 30 a.u. for plot purpose.

We further investigate the source of nonadiabaticity in aniline photodissociation: the 3 conical intersections. Figure 6 presents the geometries of mex-$^1B_1$-$^1B_2$, mex-$^1B_2$-$^1A_2$ and mex-$^1A_1$-$^1B_1$ along with their orthogonal[94] branching space[95] vectors **g** (energy difference gradient vector) and **h** (interstate coupling gradient vector). For mex-$^1B_1$-$^1B_2$, **g** is mostly benzene symmetric stretching combined with C1 out of $NH_2$ plane angle, **h** is mainly benzene torsion angle coupled with N-H asymmetric stretching, so the conical intersection anharmonicity would allow the exchange of vibrational energy between the benzene ring and the amino group. For mex-$^1B_2$-$^1A_2$ and mex-$^1A_1$-$^1B_1$, **g** and **h** are dominated by N-H stretching and C1 out of $NH_2$ plane angle.

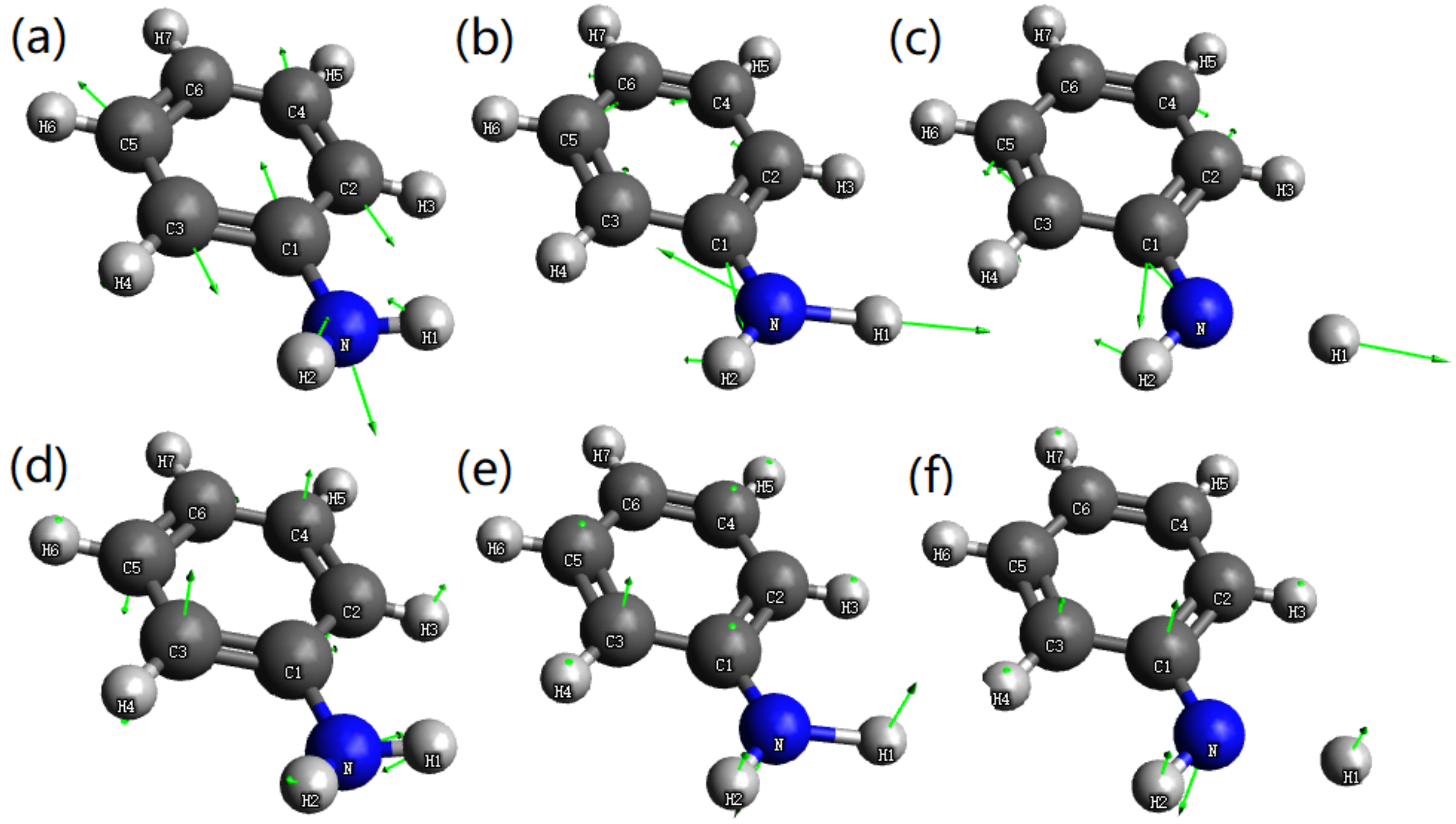

**Figure 6.** Aniline (a)(d) $^1B_1$-$^1B_2$, (b)(e) $^1B_2$-$^1A_2$, (c)(f) $^1A_1$-$^1B_1$ minimum energy conical intersections, along with (a)(b)(c) **g**, the energy difference gradient vector; (d)(e)(f) **h**, the interstate coupling gradient vector.

Figure 7 reports the double cone topography of the 3 conical intersections in **g**-**h** plane. Instead of a common double cone, the $^1B_1$-$^1B_2$ surfaces form a dual double cone. The $^1B_2$-$^1A_2$ and the $^1A_1$-$^1B_1$ double cones are more standard.

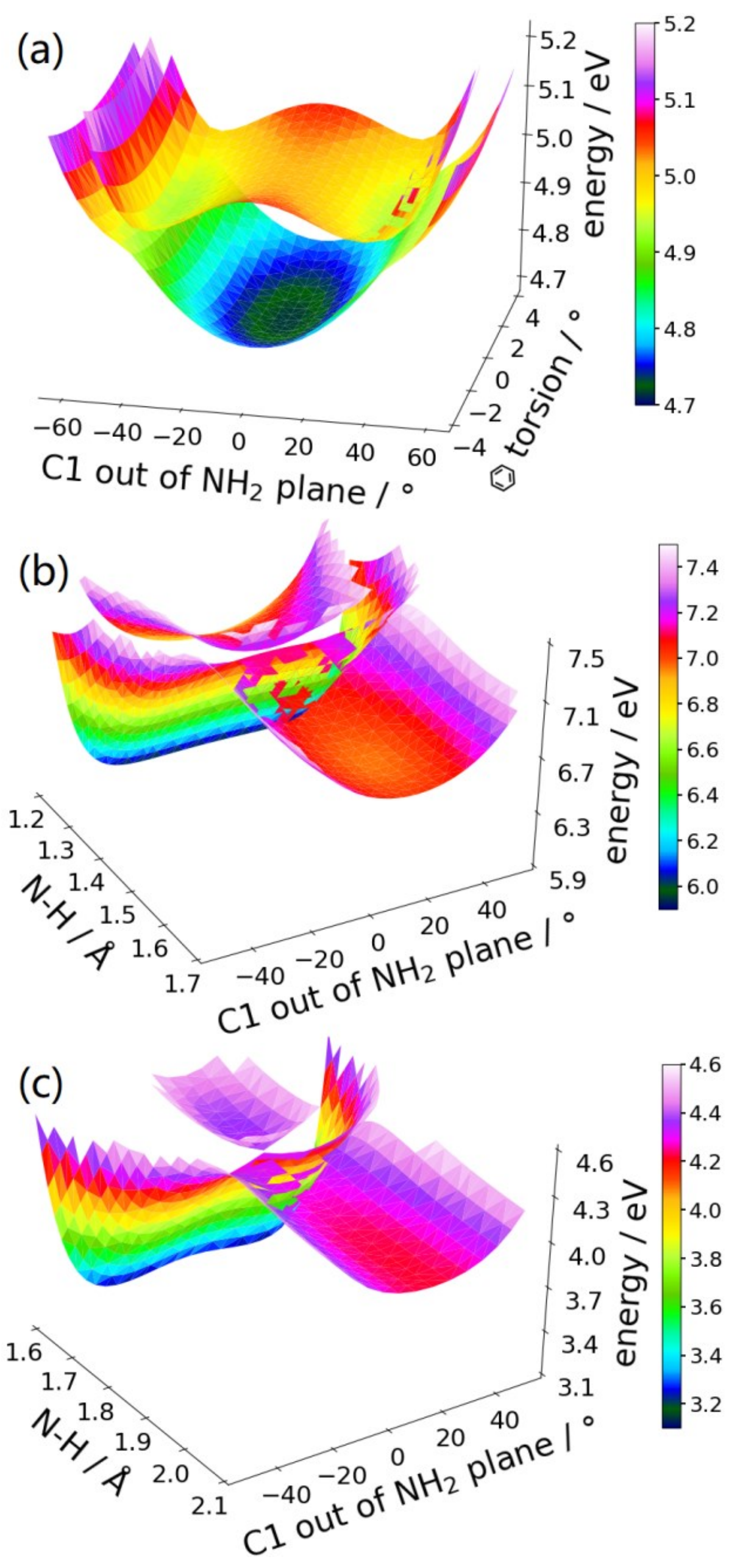

**Figure 7.** (a) The dual double cone topography formed by the $^1B_1$-$^1B_2$ surfaces around the dual $^1B_1$-$^1B_2$ minimum energy conical intersections. (b) The double cone topography formed by the $^1B_2$-$^1A_2$ surfaces around the $^1B_2$-$^1A_2$ minimum energy conical intersection. (c) The double cone

topography formed by the $^1A_1$-$^1B_1$ surfaces around the $^1A_1$-$^1B_1$ minimum energy conical intersection.

The geometric phase effect[96] is one of the important consequences of conical intersections: the loop integral of the nonadiabatic coupling ($\vec{d}$) around a conical intersection is ±π, which means that when nuclei loop around a conical intersection, in adiabatic representation electrons would not restore their states, instead they have accumulated a phase. When multiple conical intersections are linked, an induced geometric phase effect[97] would occur: the loop integral is no longer necessarily ±π, and the loop integral can even have a dependence on the starting location. For aniline, one such illuminating example is the loop around the dual $^1B_1$-$^1B_2$ crossings and the $^1B_2$-$^1A_2$ crossing, which is presented in Figure 8.

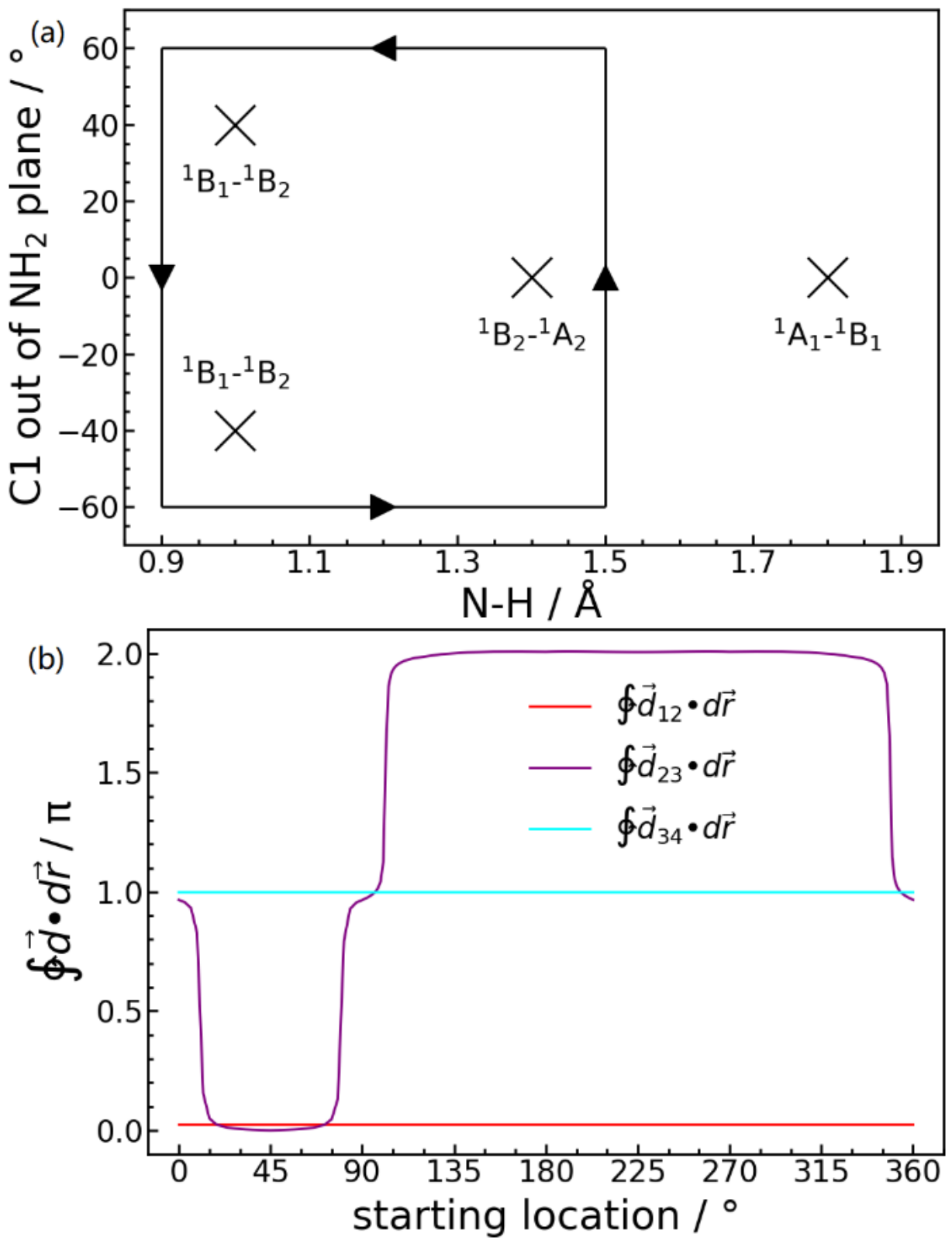

**Figure 8.** (a) The integration loop. (b) The loop integral of the nonadiabatic coupling ($\vec{d}$) as a function of the starting location, with the lower left corner assigned as 0°.

$\oint \vec{d}_{23} \cdot d\vec{r}$ demonstrates a strong induced geometric phase effect: it greatly deviates from 2π (the expected value if the $^1B_2$-$^1A_2$ crossing was absent), and a severe starting location dependence emerges. In another word, when nuclei loop as Figure 8a, depending on the starting location, the adiabatic electronic wave functions can change in different ways as Figure 8b shows. Recall the trajectory in Figure 3, the nuclei would spend hundreds of femtoseconds around the dual $^1B_1$-$^1B_2$ crossings, which suggests that the geometric phase and the induced geometric phase effects would contribute to the photodissociation dynamics. In addition, the role of mex-$^1B_2$-$^1A_2$ has been revealed: although it does not involve directly in the photodissociation dynamics since the

$^1B_2$ and the $^1A_2$ states are not populated upon reaching it, it indirectly participates through the induced geometric phase effect.

## IV. CONCLUSION

We have presented a SAResNet diabatization method to construct $\boldsymbol{H}^{\mathrm{d}}$ that accurately represents *ab initio* adiabatic energies, energy gradients, and nonadiabatic couplings for moderate sized systems. Our SANN inherits from the pioneering SAP and FINN diabatization methods to exploit the power of NN along with the transparent symmetry adaptation of polynomial for both symmetric and asymmetric irreducible representations. In addition, our symmetry adaptation provides a unified framework for SAP and SANN, enabling the adoption of ResNet architecture, which is a powerful descendant of the pioneering feedforward NN. Our SAResNet is applied to construct the full 36-dimensional coupled diabatic potential energy surfaces for aniline N-H bond photodissociation, with 2,269 data points and 32,640 trainable parameters and 190 $cm^{-1}$ root mean square deviation in energy. In addition to the experimentally observed $^1B_1$ and $^1B_2$ states, a higher $^1A_2$ state is found to introduce an induced geometric phase effect thus indirectly participate in the photodissociation process. The photodissociation dynamics of aniline will be determined employing the constructed $\boldsymbol{H}^{\mathrm{d}}$ in future work. Intersystem crossings in aniline[98–100] will also be investigated in future work.

## ACKNOWLEDGMENTS

This work was funded by National Science Foundation (NSF), Grant No. CHE-1954723 to D.R.Y. Computer time was provided by Advanced Research Computing at Hopkins (ARCH) funded by NSF, Grant No. OAC1920103.

## SUPPORTING INFORMATION

Details of aniline computation: the critical geometries, the methodology of electronic structure calculation, the point group and complete nuclear permutation inversion group, the internal coordinate system. This information is available free of charge via the Internet at http://pubs.acs.org

## NOTES

The authors declare no competing financial interest.

## AUTHOR INFORMATION

### Corresponding Authors

**Yifan Shen** - *Department of Chemistry, Johns Hopkins University, Baltimore, Maryland 21218, United States*; orcid.org/0000-0003-0590-444X; Email: yifanshensz@jhu.edu

**David R. Yarkony** - *Department of Chemistry, Johns Hopkins University, Baltimore, Maryland 21218, United States*; orcid.org/0000-0002-5446-1350; Email: yarkony@jhu.edu

141101.

For Table of Contents Only

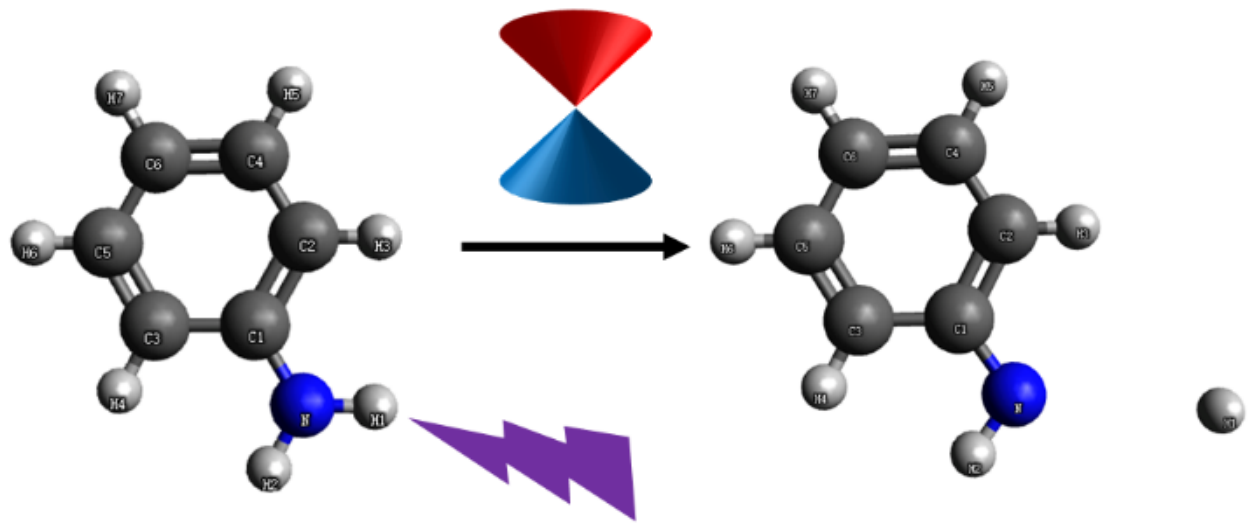